\documentclass[aps,prl,reprint,twocolumn,superscriptaddress]{revtex4}
\pdfoutput=1
\usepackage{color}

\usepackage[pdftex]{graphicx} 
\usepackage{amsmath}
\usepackage{amsfonts}
\usepackage{mathrsfs}
\usepackage{bm}
\usepackage{multirow}
\usepackage{grffile}
\usepackage{braket}

\usepackage{times}

\begin{document}

\title{Moving perturbation in a one-dimensional Fermi gas}
\author{A.-M. Visuri}
\affiliation{COMP Centre of Excellence, Department of Applied Physics, Aalto University, FI-00076 Aalto, Finland}

\author{D.-H. Kim}
\affiliation{Department of Physics and Photon Science, School of Physics and Chemistry, Gwangju Institute of Science and Technology, Gwangju 500-712, Korea}

\author{J. J. Kinnunen}
\affiliation{COMP Centre of Excellence, Department of Applied Physics, Aalto University, FI-00076 Aalto, Finland}

\author{F. Massel}
\affiliation{Department of Mathematics and Statistics, University of Helsinki, FI-00014
Helsinki, Finland}

\author{P. T\"{o}rm\"{a}}
\email{paivi.torma@aalto.fi}
\affiliation{COMP Centre of Excellence, Department of Applied Physics, Aalto University, FI-00076 Aalto, Finland}

\begin{abstract}
We simulate a balanced attractively interacting two-component Fermi gas in a one-dimensional lattice perturbed with a moving potential well or barrier. Using the time-evolving block decimation method, we study different velocities of the perturbation and distinguish two velocity regimes based on clear differences in the time evolution of particle densities and the pair correlation function. We show that, in the slow regime, the densities deform as particles are either attracted by the potential well or repelled by the barrier, and a wave front of hole or particle excitations propagates at the maximum group velocity. Simultaneously, the initial pair correlations are broken and coherence over different sites is lost. In contrast, in the fast regime, the densities are not considerably deformed and the pair correlations are preserved. 
\end{abstract}

\maketitle

In three dimensions, the superfluid phase can be broken by excitations when the fluid moves in a capillary at a velocity that is larger than the critical velocity \cite{Landau}, or by e.g.~moving an object \cite{Pickett}, a laser beam \cite{Ketterle1, Ketterle2}, or an optical lattice \cite{MovingLattice} through the superfluid at a high enough velocity. In a recent experiment, a laser beam was rotated in a two-dimensional quasi-condensate to find the critical velocity of a BKT transition \cite{Dalibard}. In this study, we simulate a perturbation propagating in a one-dimensional (1D) lattice and find that the initial pair-correlated state is, in contrast to higher-dimensional systems, broken by a perturbation with velocity {\it{below}} a certain limit. According to Landau's criterion, elementary excitations with energy $\epsilon(q)$ and momentum $q$ can appear in a superfluid if the velocity of the superfluid with respect to the capillary is larger than the critical velocity \cite{Landau}, $ v > v_{\text{c}} = \min_q \frac{\epsilon(q)}{|q|}$. In a Fermi superfluid in two or three dimensions, the single-particle (BCS) dispersion relation is $E(k) = \sqrt{(k^2 - \mu)^2 + \Delta^2}$, and the elementary excitations are particle-hole excitations close to the Fermi surface with energy $\epsilon(q) = E(\pm k_F) + E(\pm k_F + q)$. The minimum of $\frac{\epsilon(q)}{|q|}$ is found at $q = \mp 2k_F$, which gives the critical velocity $\frac{\Delta}{k_F}$ for the excitation of a quasiparticle pair. For Bose-Einstein condensates, experiments have shown that weak perturbations break the superfluidity by creating phonon excitations~\cite{Moritz} and strong perturbations by vortices~\cite{Ketterle1, Ketterle2, Anderson}. A recovery of superfluidity at high velocities of a perturbing laser beam has also been observed \cite{Atherton}.

Collective excitations in a Fermi liquid can decay into the constituent quasiparticle excitations due to the continuum of low-energy states. In one dimension, there are no zero-energy excitations with momentum transfer $0 < q < 2 k_F$, and collective excitations remain stable \cite{Voit}. In the Luttinger liquid model, the dispersion relation is linearized at the Fermi momentum $k_F$ and the slope gives the velocity of long-wavelength collective excitations (sound waves). In an interacting two-component Fermi gas, the spin and charge excitations propagate at different velocities denoted by $u_{\sigma}$ and $u_{\rho}$ \cite{Giamarchi}. For attractive interactions, the long-wavelength properties are described by $u_{\rho}$ and the exponent of the power law decaying correlation functions $K_{\rho}$. The speed of sound $v_{\text{s}}$ is equal to the velocity of charge excitations $u_{\rho}$, which, for the Hubbard model of interest here, can be solved numerically for any interaction from the Bethe Ansatz (BA).

One might expect to excite sound waves by perturbing the system. To model the critical velocity experiments, we use wave-packet perturbations which are not localized in momentum or frequency space, and do not excite a specific mode but a collection of modes. Therefore, modes with velocity higher than $v_{\text{s}}$ can also be excited. The maximum group velocity $v_{\text{g}}^{\text{max}} = \frac{d E(k)}{dk}|_{k = \frac{\pi}{2}}$ can be calculated from the lattice dispersion in the limiting cases of a non-interacting system $U = 0$ and strong interactions $|U| \gg J$. The free-particle dispersion relation in a homogeneous lattice is $E(k) = -2 J \cos{k}$, and in the strong coupling limit, the Hamiltonian is mapped to an isotropic Heisenberg Hamiltonian \cite{1DHubbardModel} and the doublons propagate as hard-core bosons with $E(k) = \frac{4 J^2}{U} \cos{k}$. The values of $v_{\text{g}}^{\text{max}}$ together with the values of $u_{\rho}$ are given in Table~\ref{table:table1} for different interactions $U$. It is of interest to study velocities of the perturbation above and below these values.

\begin{table}[!h]
\caption{Velocities of the density wave fronts $v_{\text{w. f.}}^{\uparrow}$ at different values of $U$ and the velocity of the perturbation $v$ obtained from the simulations (see Supplemental Material) \cite{supplemental}, with errors below 0.01 $J$. These are close to the maximum group velocities $v_{\text{g}}^{\text{max}}$ calculated from the lattice dispersion in the non-interacting limit for $U = 0$ and in the strong-coupling limit for $U \leq -3 \: J$. We also quote the BA results for $u_{\rho}$ (with error 0.1 $J$) solved for a homogeneous system with uniform density \cite{Giamarchi_Shastry}, using the average density between lattice sites 25 and 75.}
\centering
\begin{tabular}{c | c | c | c | c | c}
				    &$U$ ($J$)	&$v$ ($J$)	&$v_{\text{w. f.}}^{\uparrow}$ ($J$)	 &$v_{\text{g}}^{\text{max}}$ ($J$)	&$u_{\rho}$ ($J$)	\\
\hline
\multirow{4}{*}{\shortstack{Gaussian \\perturbation, \\$V_0 = -2 \: J$}}&0	&0.5	&1.54	&2	&1.9			\\
\cline{2-6}
					&\multirow{2}{*}{-4}	&0.2		&0.92	&\multirow{2}{*}{1.0}		&\multirow{2}{*}{1.0}	\\
					&						&0.5		&0.94	&							&						\\
\cline{2-6}
					&-10					&0.2		&0.53	&0.4						&0.4					\\
\hline
\multirow{8}{*}{\shortstack{Lorentzian \\perturbation, \\$V_0 = 2 \: J$}}&\multirow{2}{*}{-3}&0.2&1.19 	&\multirow{2}{*}{1.3}	&\multirow{2}{*}{1.2}	\\
					&						&0.5		&1.30	&							&						\\
\cline{2-6}
					&\multirow{2}{*}{-4}	&0.2		&1.02	&\multirow{2}{*}{1.0}		&\multirow{2}{*}{1.0}	\\
					&						&0.5		&1.14	&							&						\\
\cline{2-6}
					&\multirow{2}{*}{-5}	&0.2		&0.83	&\multirow{2}{*}{0.8}		&\multirow{2}{*}{0.8}	\\
					&						&0.5		&1.01	&							&						\\
\cline{2-6}
					&\multirow{2}{*}{-6}	&0.2		&0.80	&\multirow{2}{*}{0.7}		&\multirow{2}{*}{0.7}	\\
					&						&0.5		&0.90	&							&						\\
\hline
\end{tabular}
\label{table:table1}
\end{table}

The time-evolving block decimation (TEBD) method \cite{Vidal, Daley} is used for calculating the ground state properties of the attractive Fermi-Hubbard Hamiltonian, including a trap to model a potential realization in ultracold gases,
\begin{align}
 H_0 = -J \sum_{i, \sigma} c_{i\sigma}^{\dagger} c_{i + 1 \sigma} + h.c. + H_U + H_{\text{trap}}.
    \label{eq:hamiltonian}
\end{align}
The terms are $H_U = U \sum_i \hat{n}_{i\uparrow} \hat{n}_{i\downarrow}$, and $H_{\text{trap}} = V_{\text{trap}}\sum_{i, \sigma} (i - C)^2 \hat{n}_{i\sigma}$, where $C$ denotes the center of the lattice. Here, $J$ is the tunneling energy, $U$ the on-site interaction energy and $V_{\text{trap}} = 5 \cdot 10^{-4} \: J$ the trapping potential in units of $J$. The particle number operator is $\hat{n}_{i\sigma} = c_{i\sigma}^{\dagger} c_{i\sigma}$, and $c_{i\sigma}$ annihilates a fermion with spin $\sigma = \uparrow, \downarrow$ on site $i = 1, \cdots, L$. The number of lattice sites is $L = 100$ and the numbers of up and down spins $N_{\sigma} = 20$. We use a Schmidt number 100 in the TEBD truncation and a time step 0.02 $\frac{1}{J}$ in the real time evolution. The results were benchmarked with earlier calculations \cite{Kollath, Heidrich-Meisner1}. TEBD and t-DMRG have been recently applied to simulating also dynamics, e.g.~in sudden expansion \cite{Heidrich-Meisner2} or in connection to impurity studies \cite{Zoller, Jaksch, Massel, Giamarchi_impurity} that are already within reach of ultracold gas experiments \cite{Bloch, Inguscio}. In the real time evolution, a perturbing potential is added and 
\begin{align}
 H(t) = H_0 + H_V(t),
 \label{eq:hamiltonian_te}
\end{align}
where $H_V(t) = \sum_{i, \sigma} V(i, t) \hat{n}_{i\sigma}$. The potential is either a Gaussian well $V(i, t) = V_0 e^{- \frac{(i - vt)^2}{2\sigma^2}}$ with $\sigma^2 = 10$ or a Lorentzian barrier $V(i, t) = \frac{\gamma}{(i - vt)^2 + \gamma^2}$, where $\gamma = \frac{1}{V_0}$, $V_0$ is the height of the potential, and $v$ is the constant propagation velocity of the perturbation. The Fourier transforms $\tilde{V}(k, \omega)$ are given in the Supplemental Material \cite{supplemental}. The exact functional form of the perturbing potential does not signify in these calculations as long as its width is small compared to the size of the lattice. Such a local perturbation leads to different physics from e.g. an accelerating optical lattice which would correspond to a vector potential \cite{Niu}.

\begin{figure}[!h]
\begin{center}
\hspace{3mm}
\includegraphics[height=0.055\linewidth]{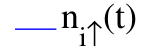}
\hspace{3mm}
\includegraphics[height=0.055\linewidth]{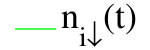}
\hspace{3mm}
\includegraphics[height=0.055\linewidth]{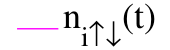}
\hspace{4mm}
\includegraphics[height=0.06\linewidth]{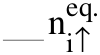}

\includegraphics[width=0.5\linewidth]{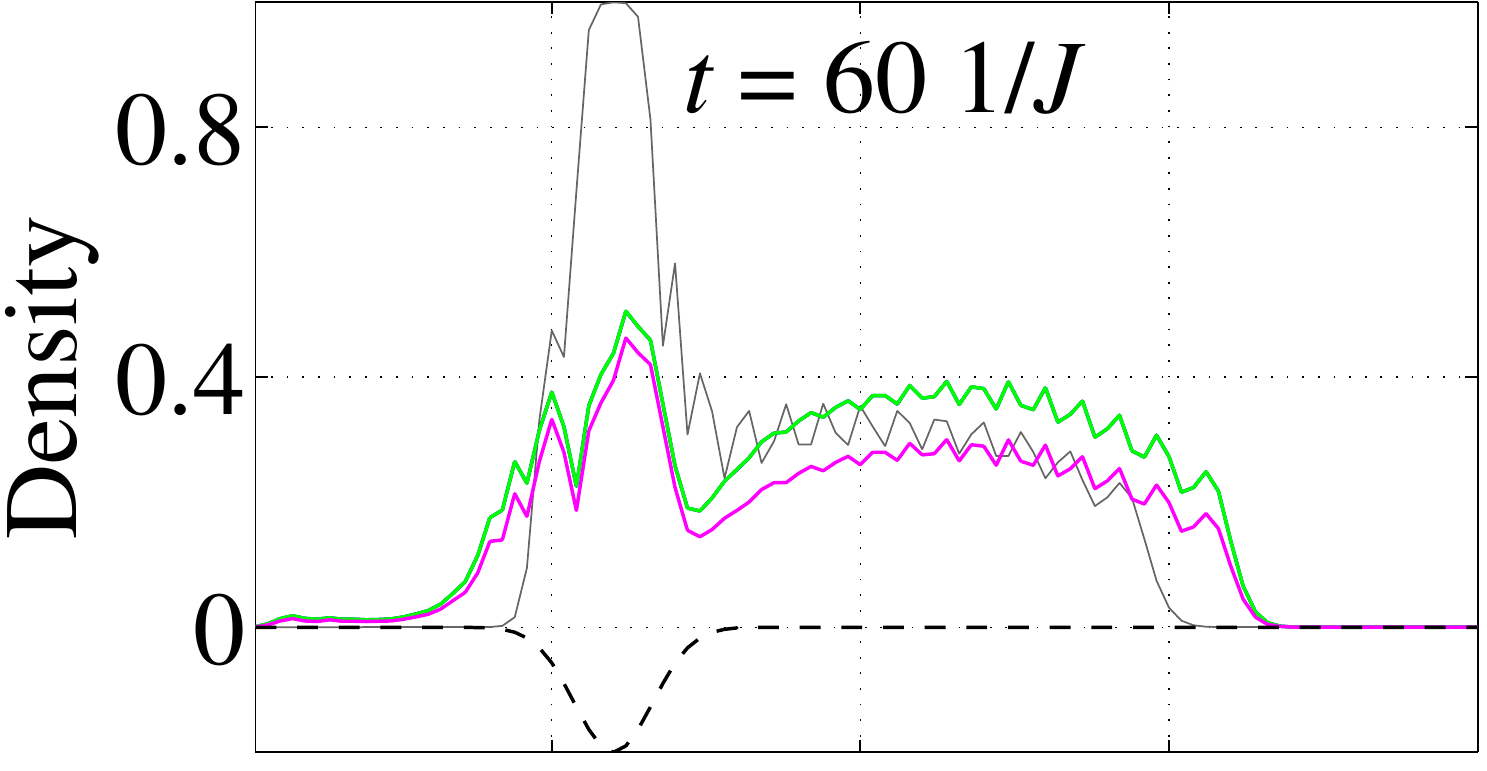}
\hspace{-2mm}
\includegraphics[width=0.5\linewidth]{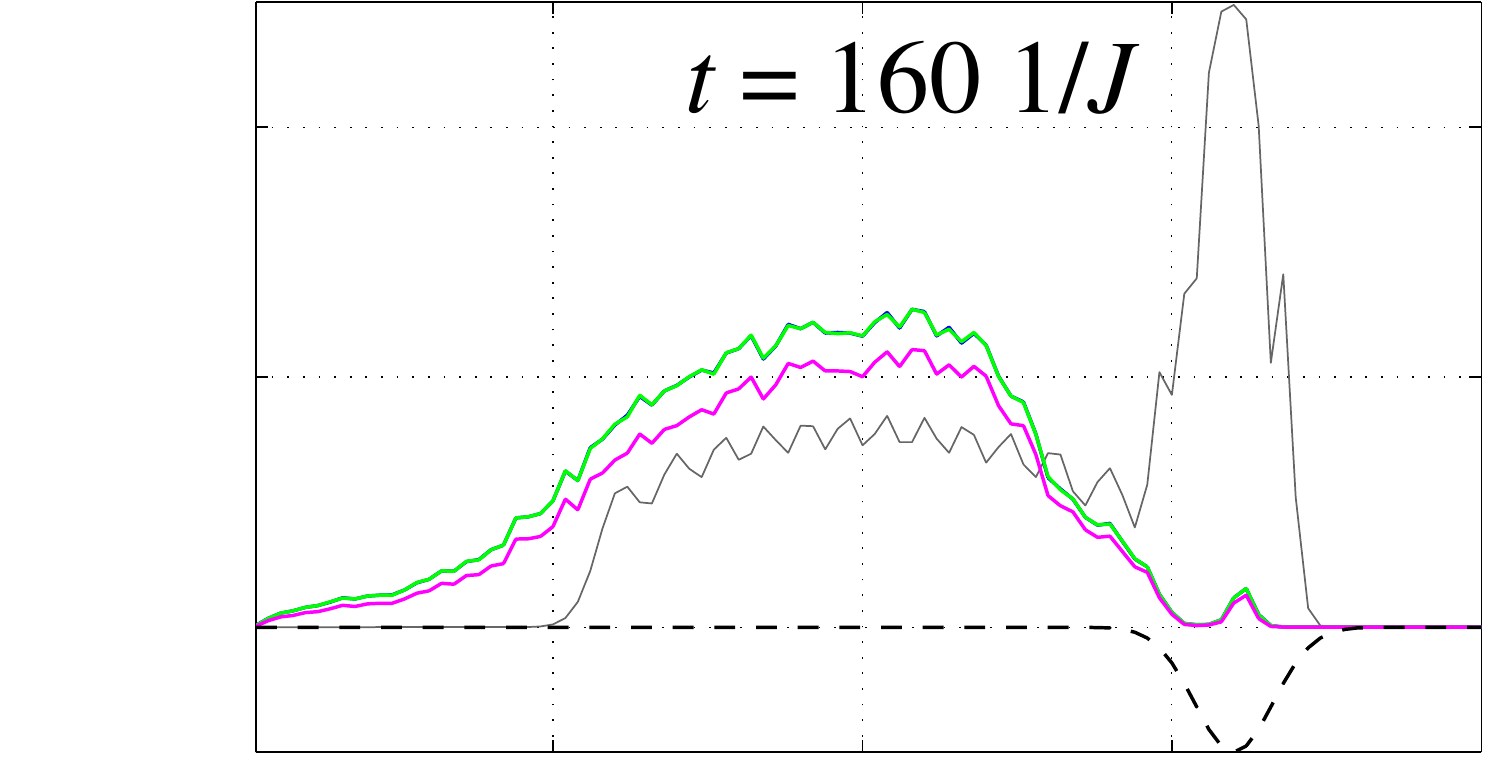}

\includegraphics[width=0.5\linewidth]{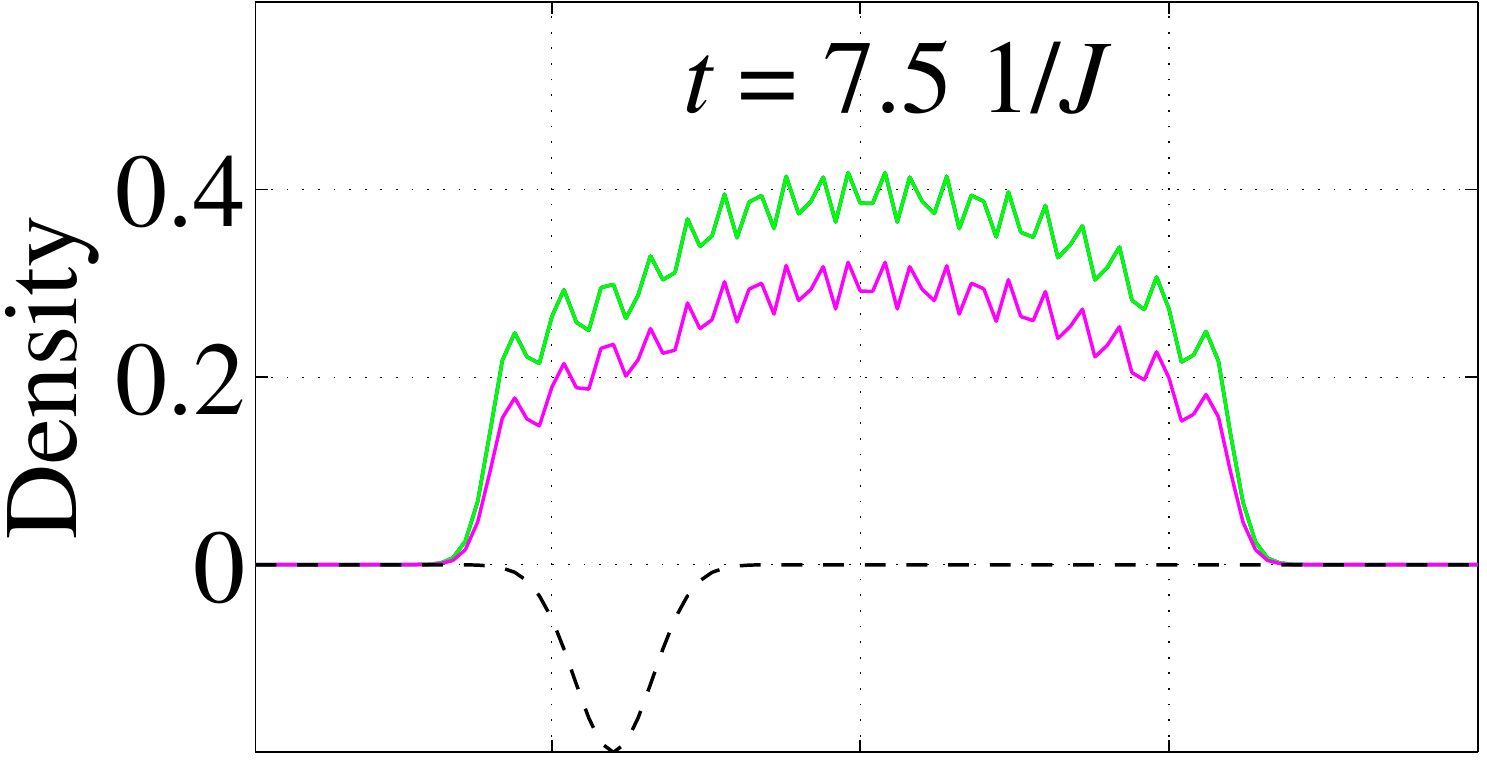}
\hspace{-2mm}
\includegraphics[width=0.5\linewidth]{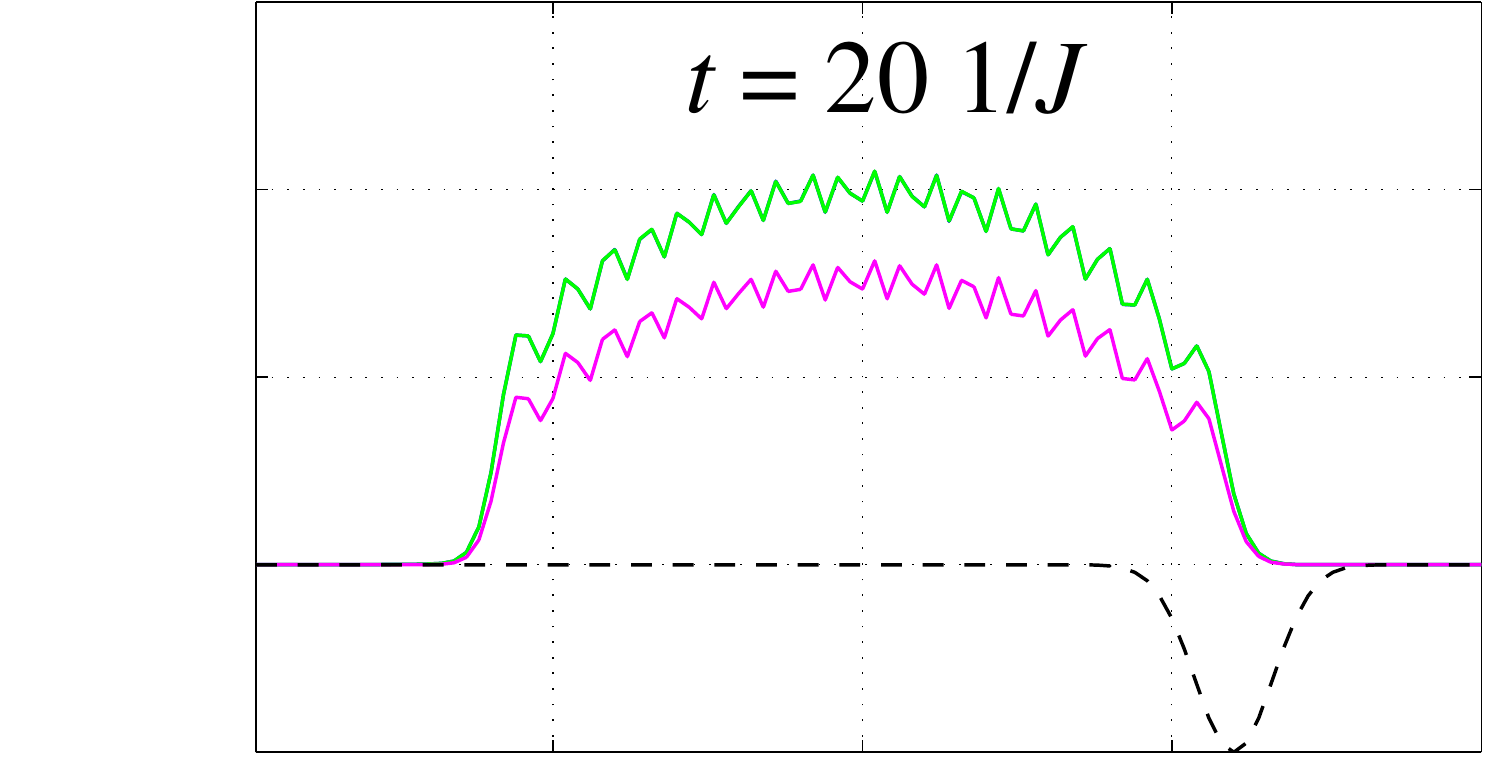}

\includegraphics[width=0.5\linewidth]{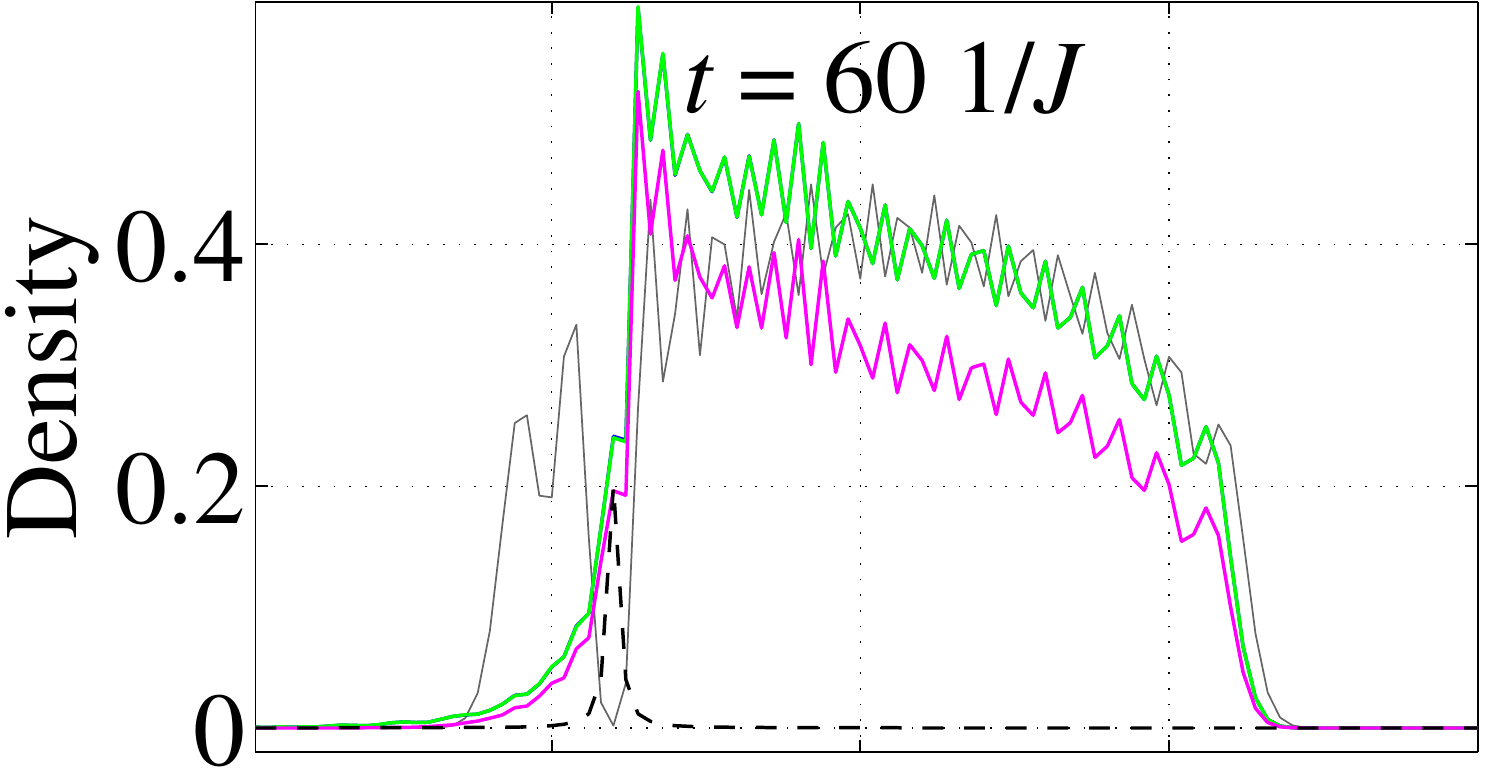}
\hspace{-2mm}
\includegraphics[width=0.5\linewidth]{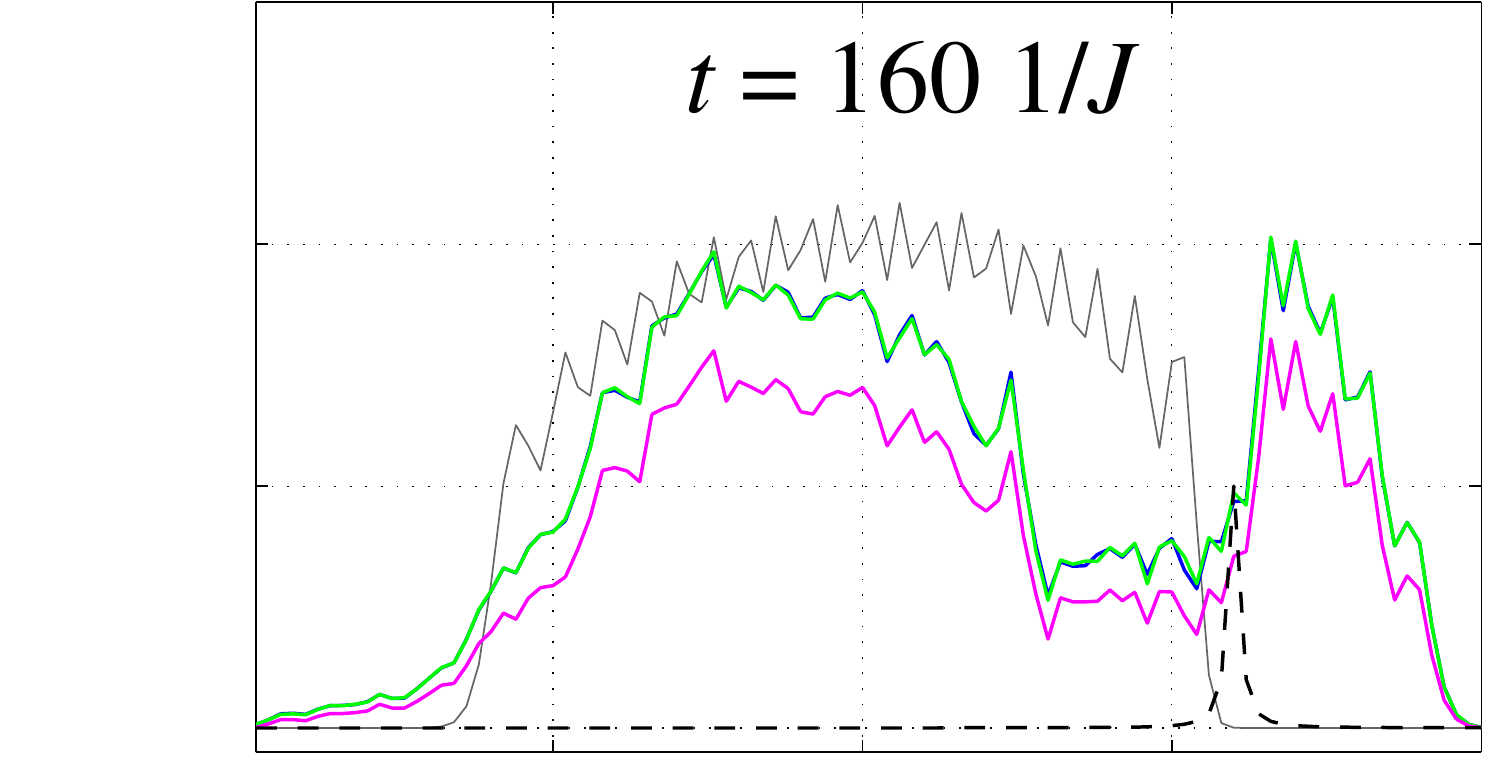}

\includegraphics[width=0.5\linewidth]{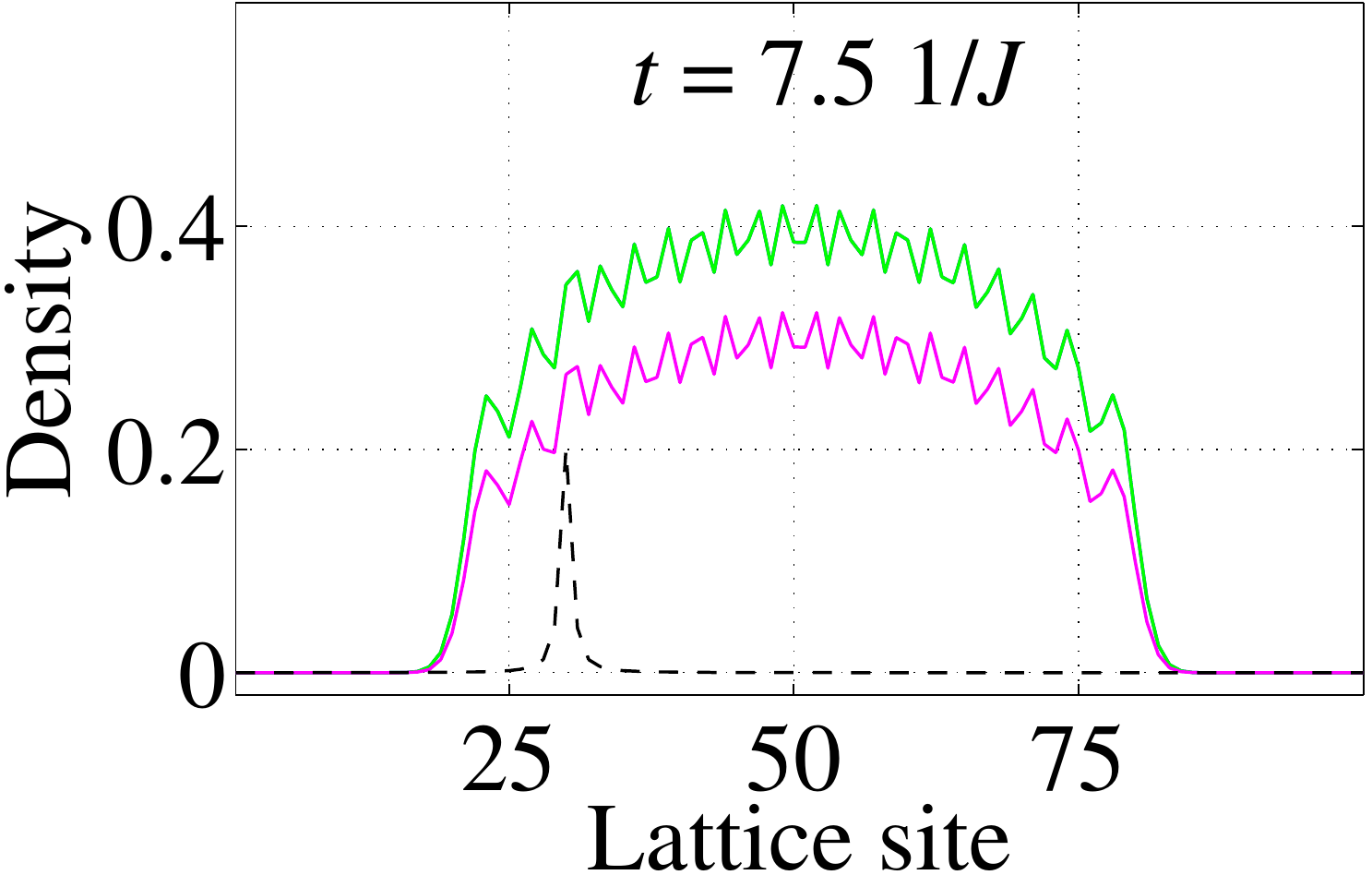}
\hspace{-2mm}
\includegraphics[width=0.5\linewidth]{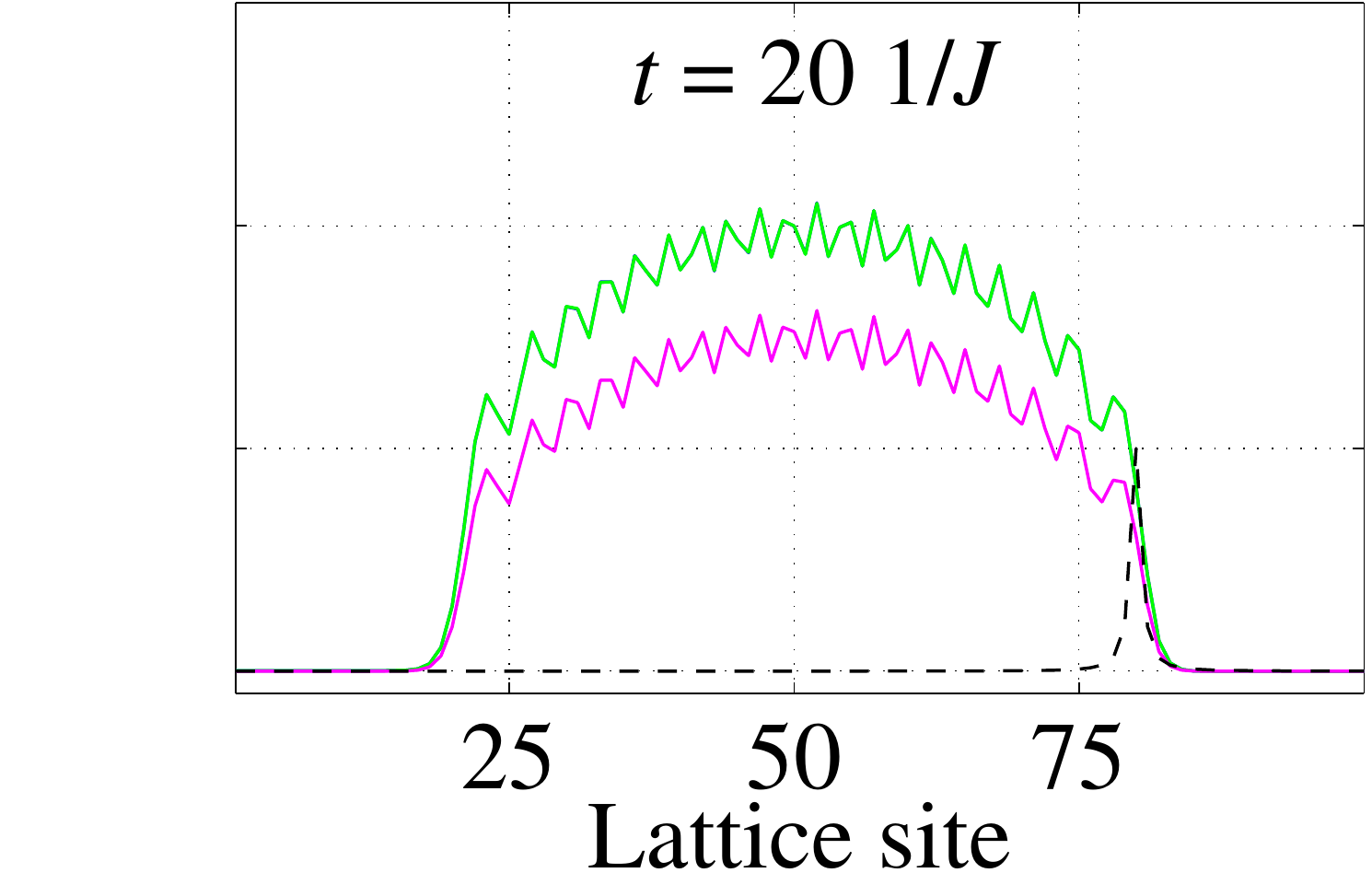}
\end{center}
\caption{(Color online). The spin-up ($n_{i\uparrow}$), spin-down ($n_{i\downarrow}$), and doublon ($n_{i\uparrow\downarrow}$) densities at times $t$ with $U = -4 \: J$ (in practice, $n_{i\uparrow}$ and $n_{i\downarrow}$ overlap). On the first and third row, the density of spin-up particles is also shown for the equilibrium case with a static potential well or barrier ($n_{i\uparrow}^{\text{eq.}}$). The first row shows a slow Gaussian well with $v = 0.5 \: J$ and the second row a fast one with $v = 4 \: J$. The third and fourth row show the same quantities in the case of a Lorentzian barrier. For the Gaussian, $V_0 = -2 \: J$ and for the Lorentzian, $V_0 = 2 \: J$. A dashed black line indicates the perturbing potential multiplied by $0.1$.}
\label{fig:densities}
\end{figure}

\begin{figure}[h]
 \begin{center}
    \includegraphics[width=0.355\linewidth]{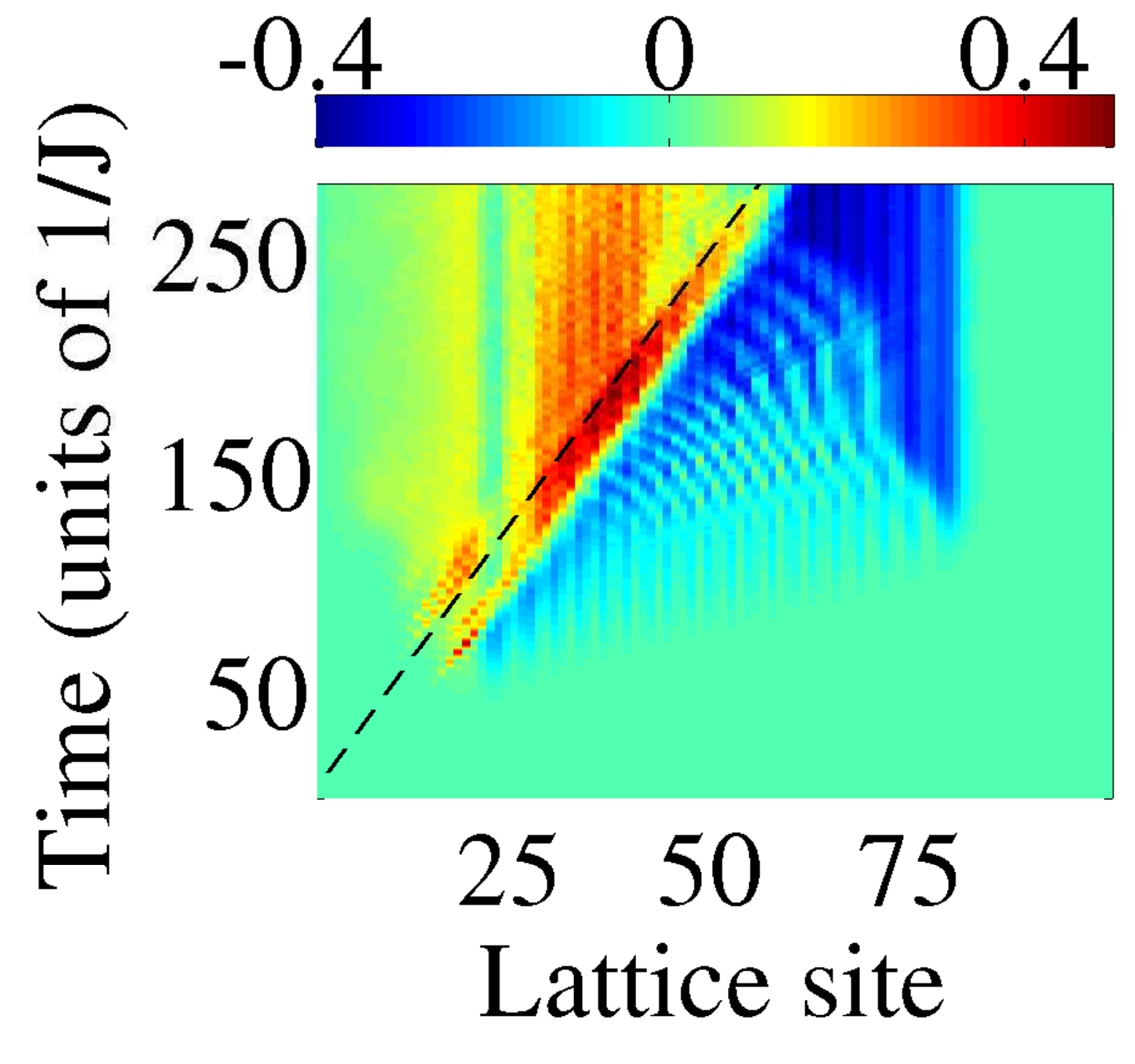}
    \includegraphics[width=0.31\linewidth]{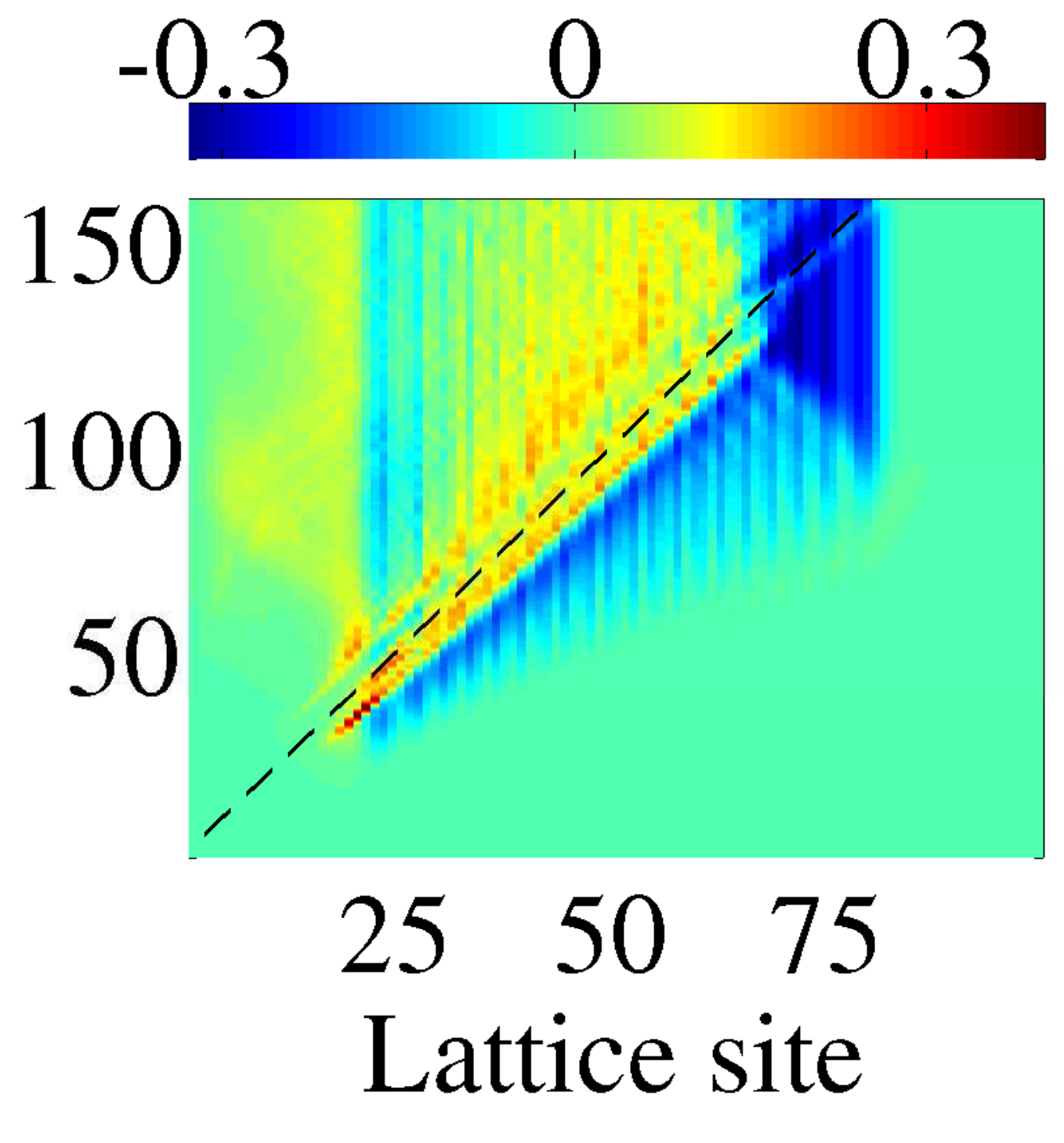}
    \includegraphics[width=0.295\linewidth]{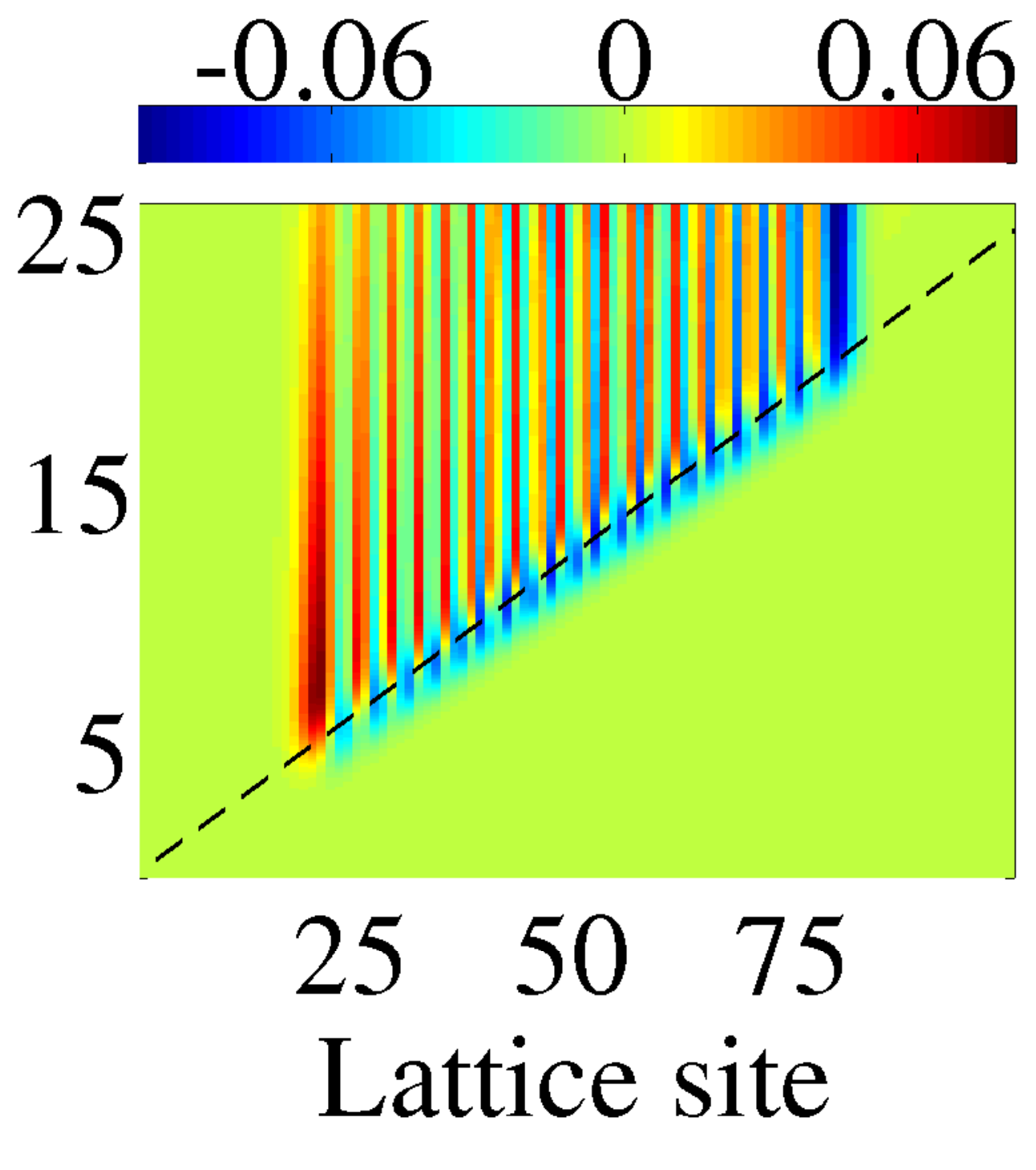}
  \caption{(Color online). The difference in the density of spin up particles with respect to the ground state as a function of position and time, $n_{i\uparrow}(t) - n_{i\uparrow}(0)$, for $U = -4 \: J$. The perturbation is a Gaussian well with $V_0 = -2 \: J$ and $v = 0.2 \: J$ (left), $v = 0.5 \: J$ (middle), $v = 4 \: J$ (right). The center of the perturbing potential is marked with a dashed black line.}
  \label{fig:density_difference_U-4-10}
 \end{center}
\end{figure}

Two approximate regimes can be distinguished in the simulation results according to the velocity of the perturbation: slow, $v \alt v_{\text{g}}^{\text{max}}$, and fast, $v \gg v_{\text{g}}^{\text{max}}$. In the slow regime, the perturbing potential produces a large deformation of the particle densities. Figure~\ref{fig:densities} shows the densities at different time steps as a Gaussian potential well or a Lorentzian barrier propagates across the lattice. The well draws in particles whereas the barrier pushes them. Comparison to the equilibrium densities for corresponding static potentials shows that the moving perturbations produce highly non-equilibrium dynamics. The movement of the particles can also be seen in Fig.~\ref{fig:density_difference_U-4-10}, which shows the density difference with respect to the ground state. For $v \alt v_{\text{g}}^{\text{max}}$, a wavefront is seen propagating faster than the perturbation and is reflected from the harmonic trap. In the case of a well, the wavefront is a reduction of density and corresponds to propagating hole excitations. For a barrier, there is an increase of density corresponding to particle excitations. The approximate velocities of the wavefronts $v_{\text{w. f.}}^{\uparrow}$ obtained from Fig.~\ref{fig:density_difference_U-4-10} and the same data for other interaction strengths are shown in Table~\ref{table:table1}. They are reasonably close to $v_{\text{g}}^{\text{max}}$ as well as the BA values $u_{\rho}$, taking into account the shallow trap. The velocity of the wavefront is independent of the velocity of the perturbation since $v_{\text{g}}^{\text{max}}$ and $u_{\rho}$ are properties of the fermion system and do not depend on $v$. The densities are perturbed less when the velocity of the perturbation is higher, as seen in Fig.~\ref{fig:densities} and in the rightmost column of Fig.~\ref{fig:density_difference_U-4-10}. There is no wave front preceding the perturbation since the velocity of the perturbation is higher than that of the excitations. The density difference that remains after the perturbation is due to the smoothening of the initial density oscillations. The oscillations indicate the tendency to singlet pairing \cite{ADW}, and their distortion in the slow regime suggests that the singlet superfluid correlations are broken. 

\begin{figure}[!h]
\begin{center}
$\vcenter{\hbox{\includegraphics[height=0.35\linewidth]{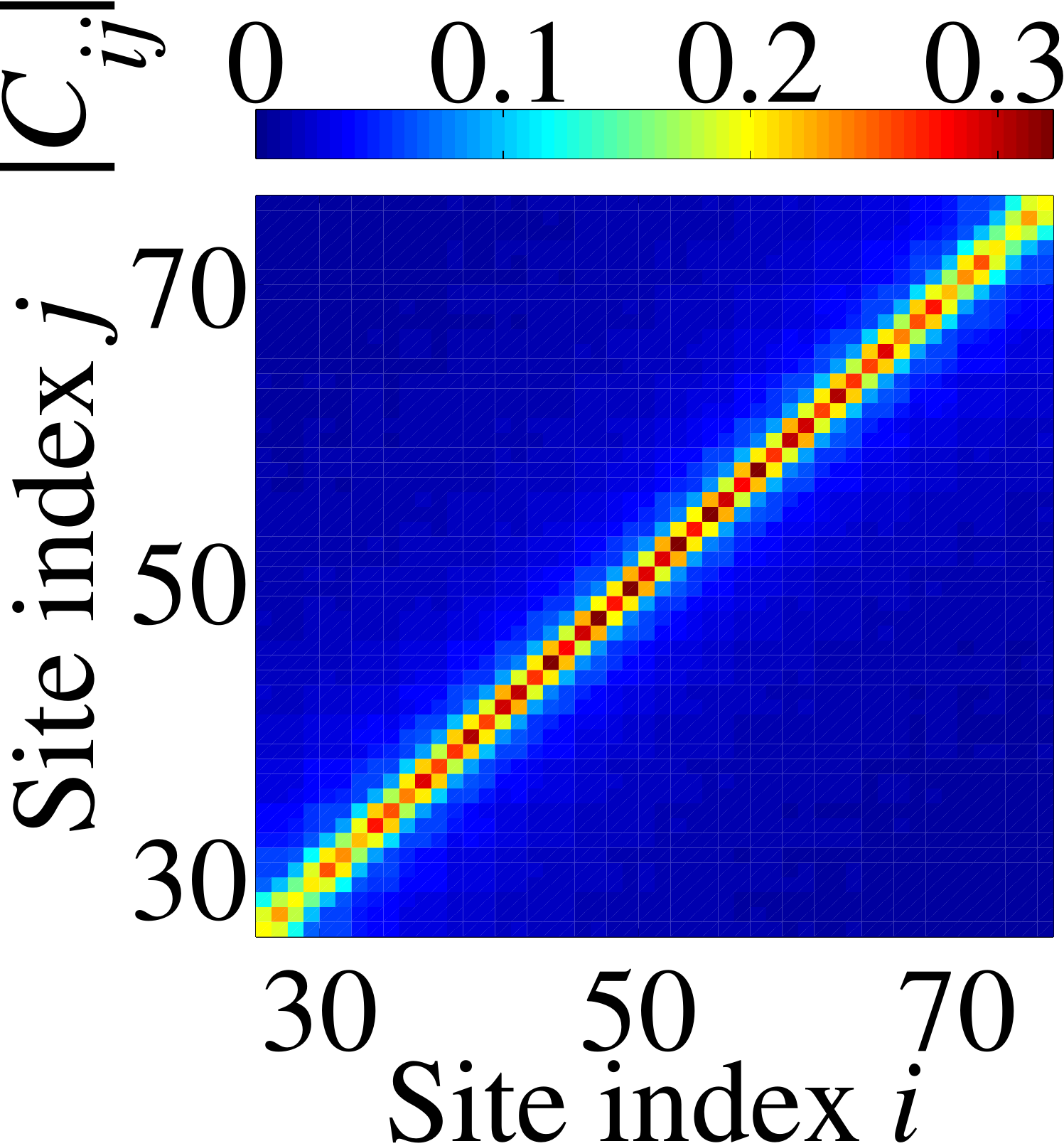}}}$
\hspace{8mm}
$\vcenter{\hbox{\includegraphics[height=0.37\linewidth]{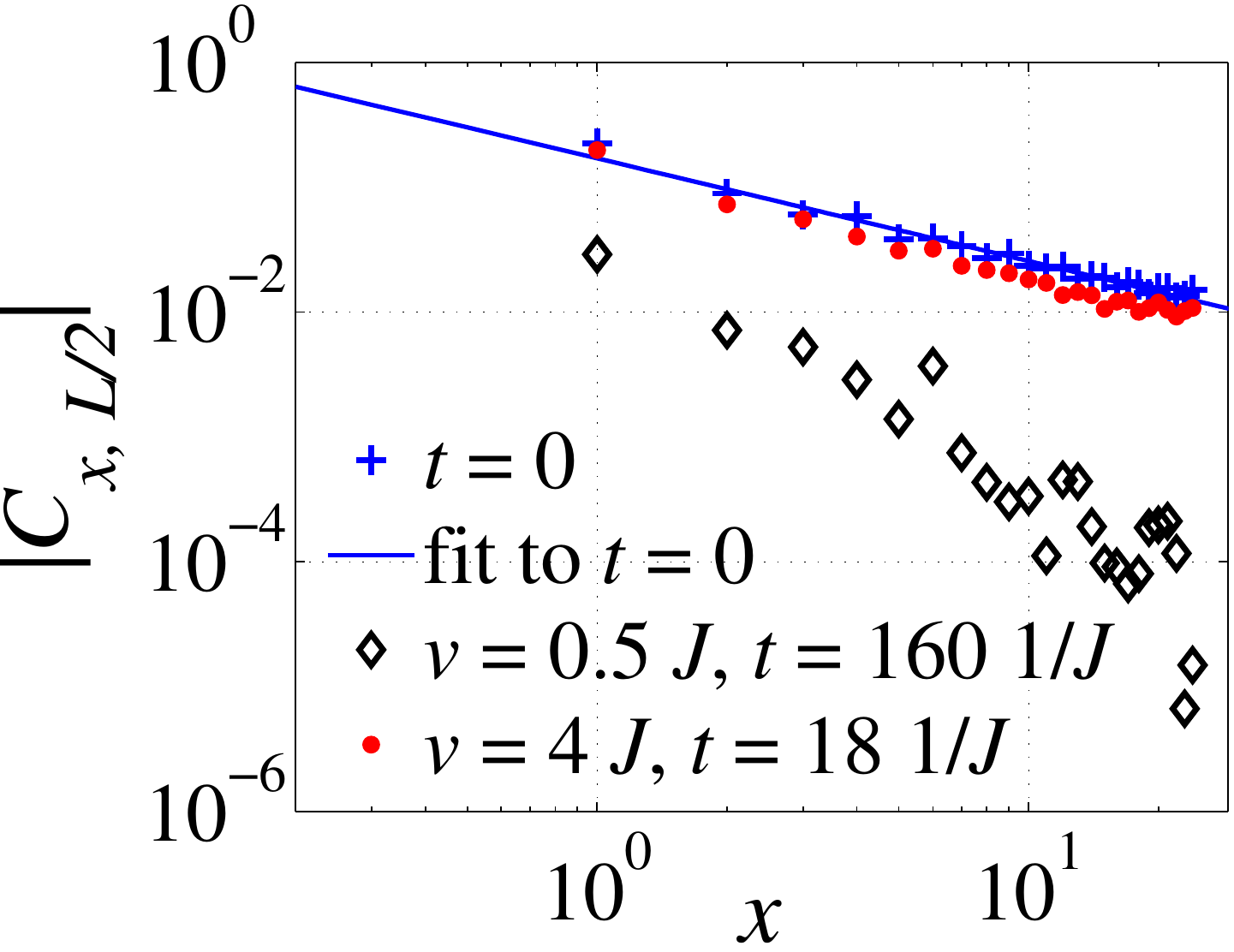}}}$
\caption{(Color online). Left: The pair correlation $|C_{ij}|$ in the ground state in the middle part of the lattice for $U = -4 \: J$. Right: The same quantity with $j$ fixed, $|C_{x, \frac{L}{2}}|$. A linear fit $f(x) = -\frac{1}{K_{\rho}}x + a$ gives the coefficients $K_{\rho} = 1.22 \pm 0.08$ and $a = -1.7 \pm 0.1$ with errors given by the 95 \% confidence bounds. Data points after moving a potential well across the center of the lattice are also shown. For a well with $v = 0.5 \: J$, $|C_{x, \frac{L}{2}}|$ is shown at the time step when the perturbation has reached the site 80 and for $v = 4 \: J$ the site 72.}
\label{fig:pair_correlation_time_evolution}
\end{center}
\end{figure}

In one dimension, there is no long-range order and the phase is determined
by the dominant power-law decaying correlation~\cite{Giamarchi}. Therefore, identifying a superfluid in 1D is not as straightforward as in higher dimensions~\cite{Castin,Leggett,Giamarchi_Shastry,Giamarchi}.  Here we study the pair correlation $C_{ij}(t) = \bra{\psi(t)} c_{i\uparrow}^{\dagger} c_{i\downarrow}^{\dagger} c_{j\downarrow} c_{j\uparrow} \ket{\psi(t)}$, which decays as $|i - j|^{-\frac{1}{K_{\rho}}}$ and contains both the off-diagonal components and the doublon density on the diagonal. This type of decay is directly connected with a nonzero spin gap~\cite{Yang, LutherEmery} and the correlator is dominant for $K_\rho > 1$, implying a singlet 1D superfluid phase for attractive interactions $U < 0$~\cite{Giamarchi, 1DHubbardModel, Feiguin}. The fit for the correlator in Fig.~\ref{fig:pair_correlation_time_evolution} gives $K_{\rho} = 1.22 \pm 0.08$.
This is close to the BA result for a homogeneous system with density 0.7, $K_{\rho} \approx 1.28 \pm 0.02$ \cite{Giamarchi_Shastry}. Figure~\ref{fig:pair_correlation_time_evolution} shows $|C_{ij}|$ in the ground state as a function of the lattice site indices $i$ and $j$. The same quantity is plotted on the right with one of the indices fixed to the center of the lattice, $|C_{x, \frac{L}{2}}|$, where $x$ is the distance from the center, in the ground state and after a time evolution with slow and fast perturbations. When applying a slowly moving perturbation, doublons move into the potential well or ahead of the barrier and lose correlations due to localization. The original many-body pairs are reduced into on-site pairs: nearly strict on-site correlations $C_{ij} \propto \delta_{ij}$ are produced instead of the initial pair correlations that extend over many lattice sites, which suggests that the 1D superfluid state is broken. Investigating properties such as the superfluid stiffness goes beyond the scope of this work.

In recent experiments, the decay of similar 1D states has been studied with nanowires~\cite{Tinkham2000a}, nanopores~\cite{Fukushima2001a, Fukushima2007a}, and oscillating atomic Bose gases~\cite{Porto2005a}. Theoretically, the onset of dissipation due to perturbations has been described by phase slips \cite{Blatter2001a,Polkovnikov2012a} or a drag force~\cite{Pavloff2002,Pitaevskii,Brand2012a} in bosonic 1D superfluids with various results depending on the interaction regime. 
Our results show that for the fermion system, the correlations are not destroyed by fast perturbations since the doublons do not have enough time to move. Only the phase of the pair correlation is shifted. A comparison to the non-interacting case reveals a dramatic difference in $|C_{ij}|$: whereas the pair correlations present in the interacting case are nearly perfectly preserved for fast velocities and destroyed for slow velocities, in the non-interacting case (see Supplemental Material) \cite{supplemental}, the decay law of correlations is practically the same for all velocities.

\begin{figure}[ht]
\includegraphics[height=0.05\linewidth]{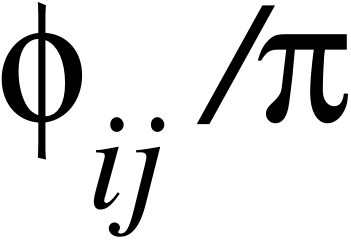}
\raisebox{1mm}{\includegraphics[height=0.05\linewidth]{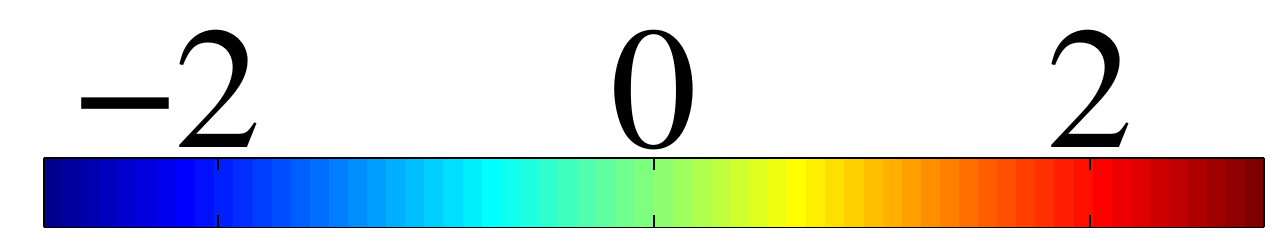}}
\hspace{0.7cm}
\includegraphics[height=0.05\linewidth]{fig4a.pdf}
\raisebox{1mm}{\includegraphics[height=0.05\linewidth]{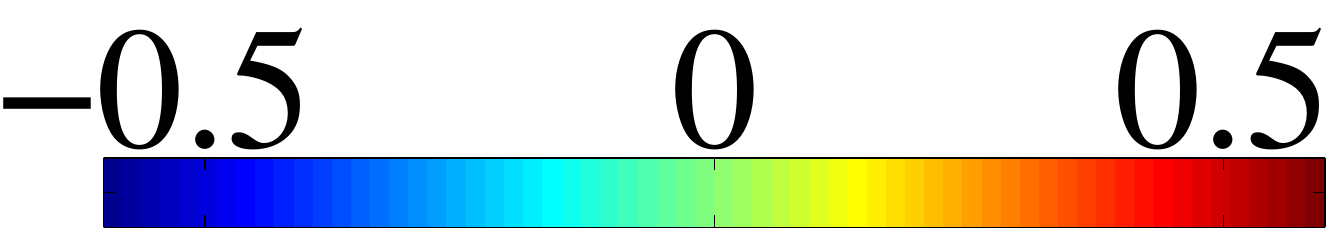}}

\centering
\includegraphics[height=0.28\linewidth]{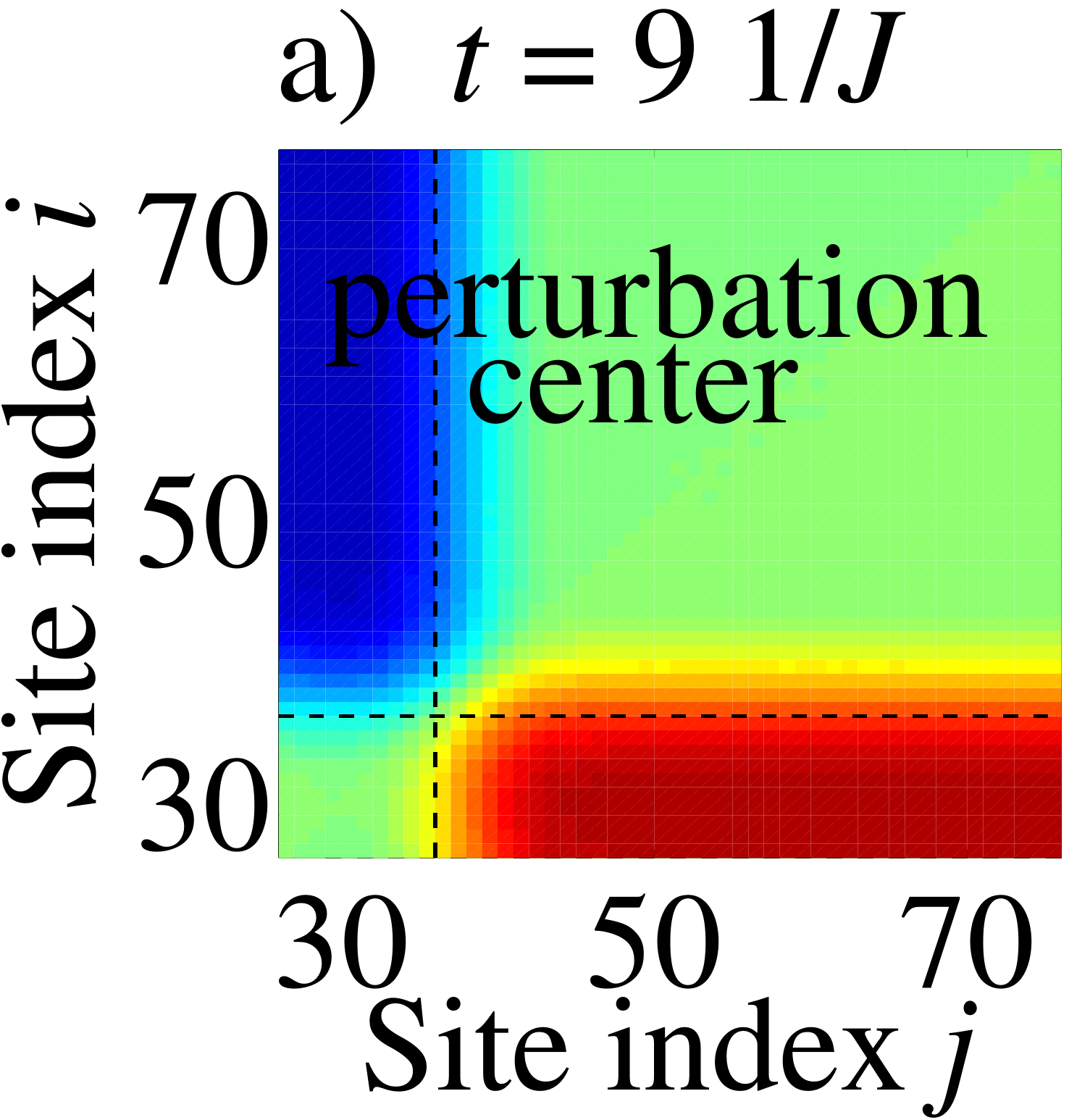}
\includegraphics[height=0.28\linewidth]{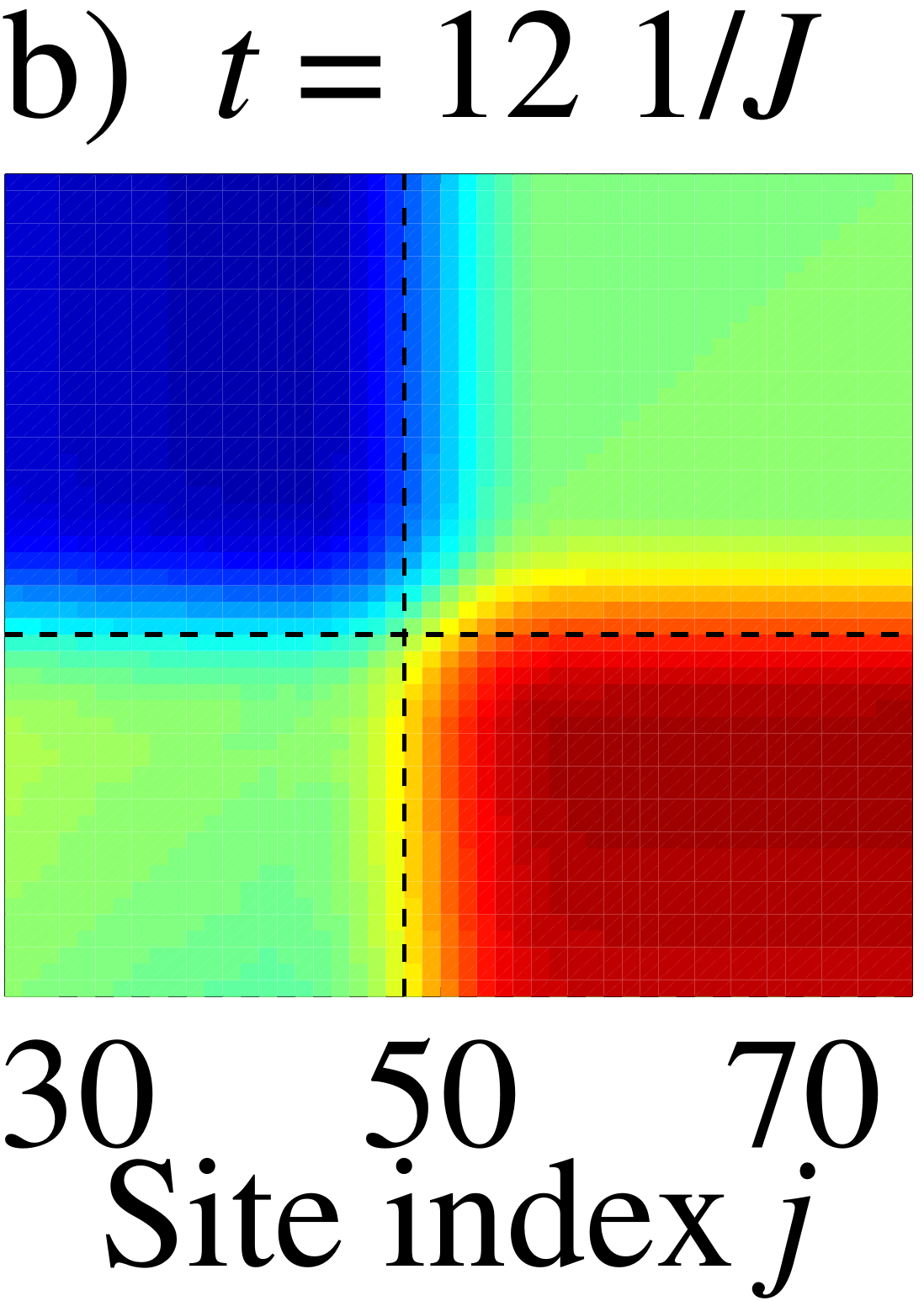}
\hspace{0.1mm}
\includegraphics[height=0.28\linewidth]{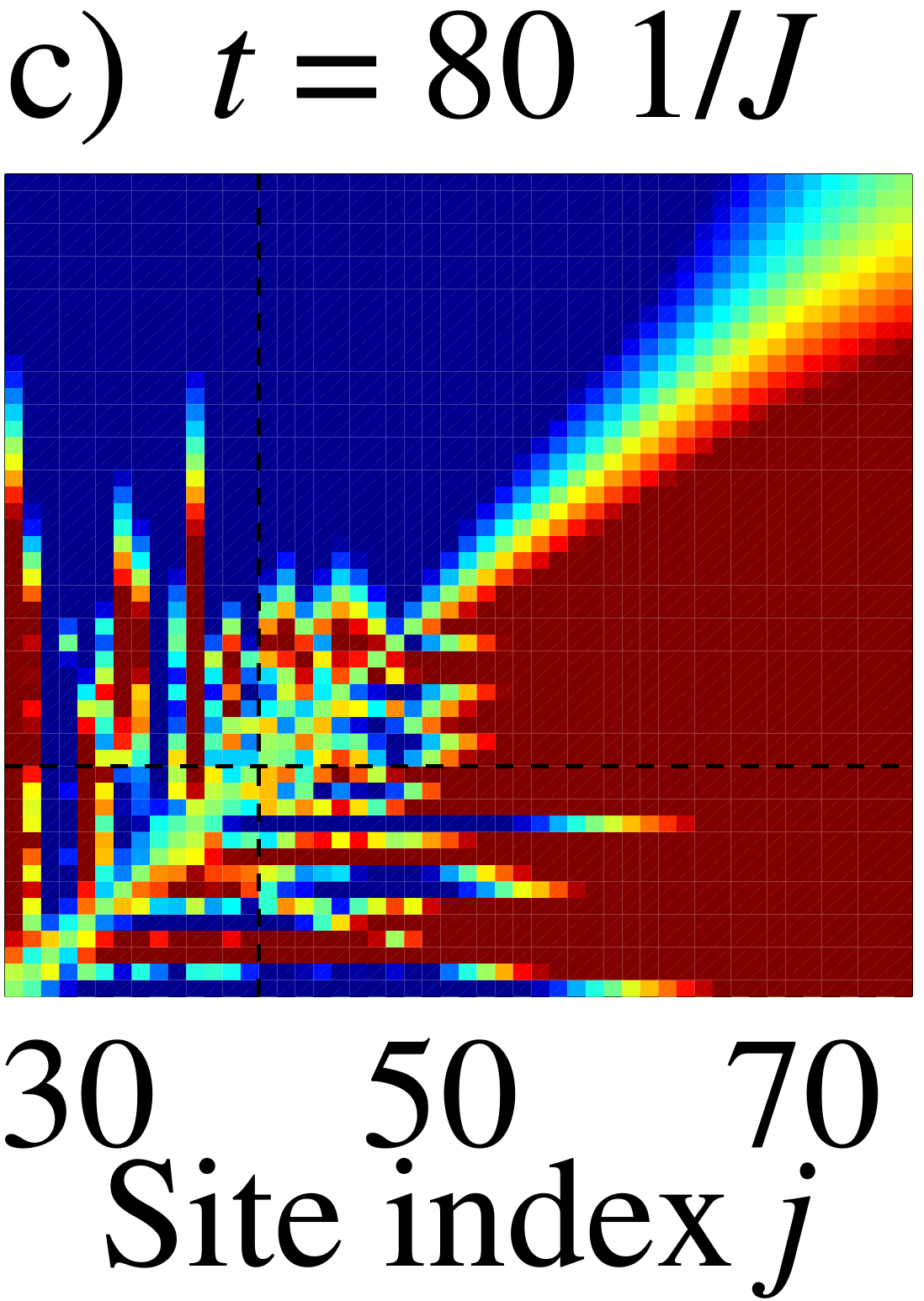}
\includegraphics[height=0.28\linewidth]{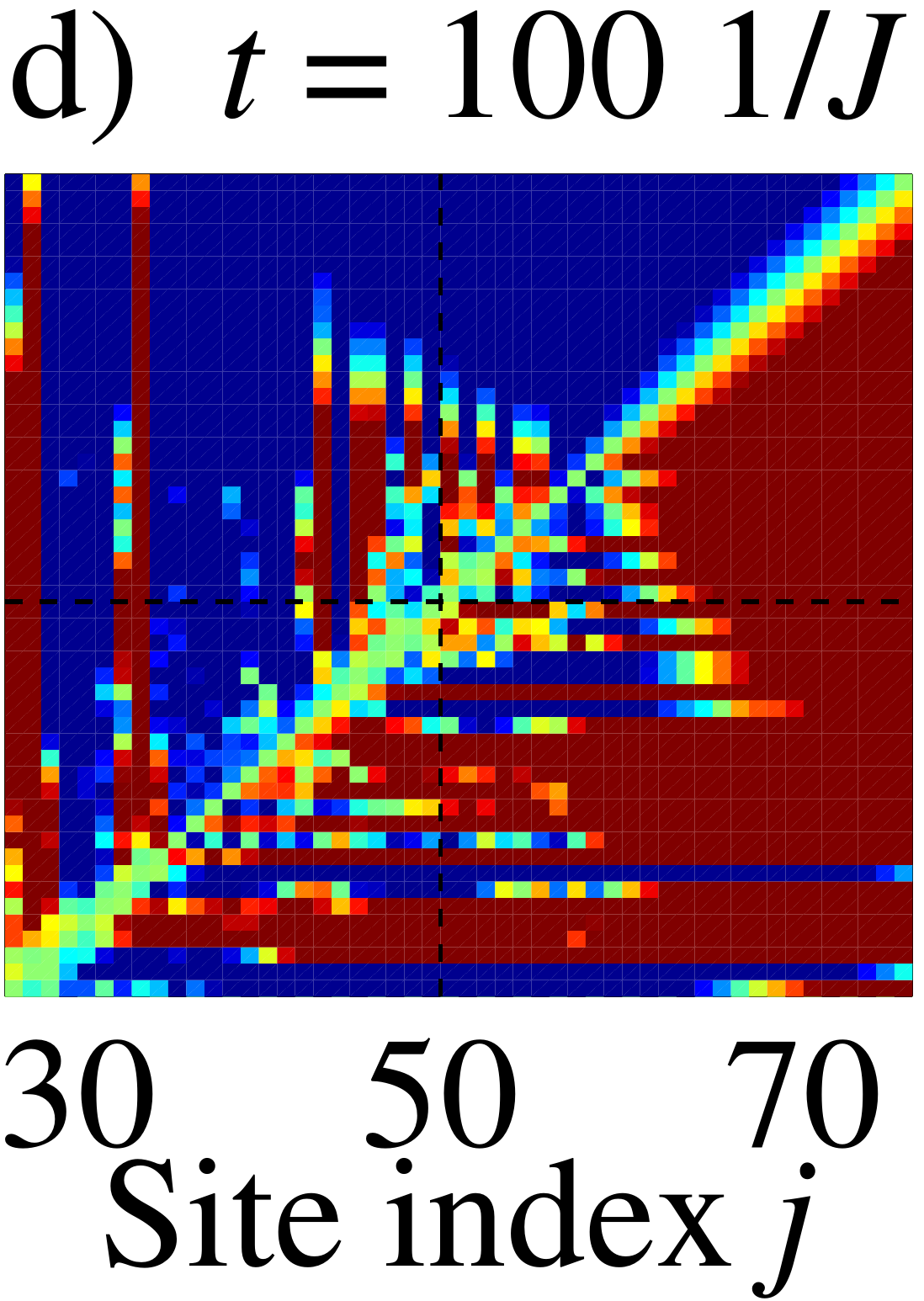}
\caption{(Color online). The phase $\phi_{ij}$ at different time steps for $v = 4 \: J$ (a, b) and $v = 0.5 \: J$ (c, d) of the Gaussian potential well with $V_0 = -2 \: J$, when $U = -4 \: J$.}
\label{fig:pair_correlation_phase}
\end{figure}

\begin{figure}[!h]
\begin{center}
$\vcenter{\hbox{\includegraphics[width=0.49\linewidth]{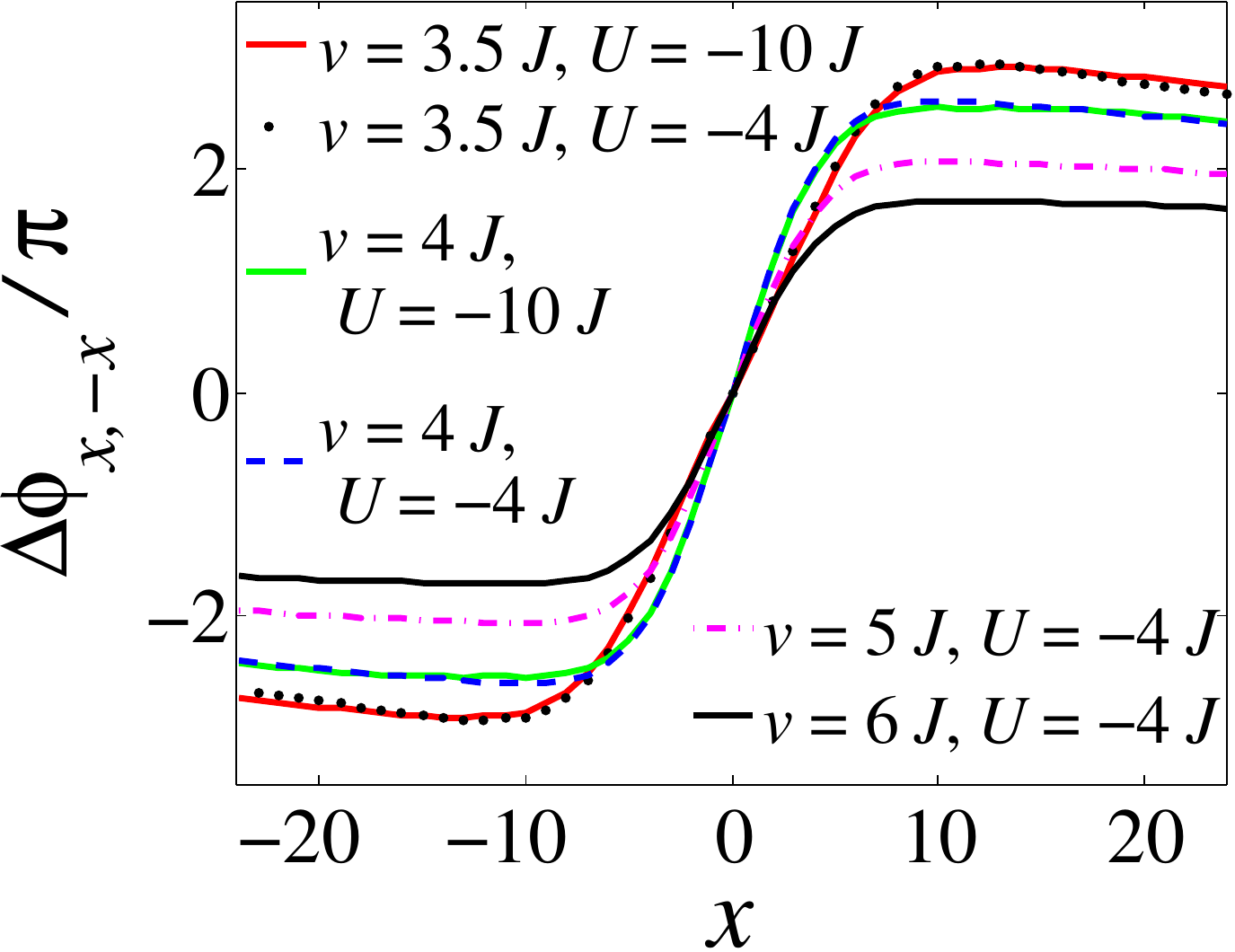}}}$
$\vcenter{\hbox{\includegraphics[width=0.49\linewidth]{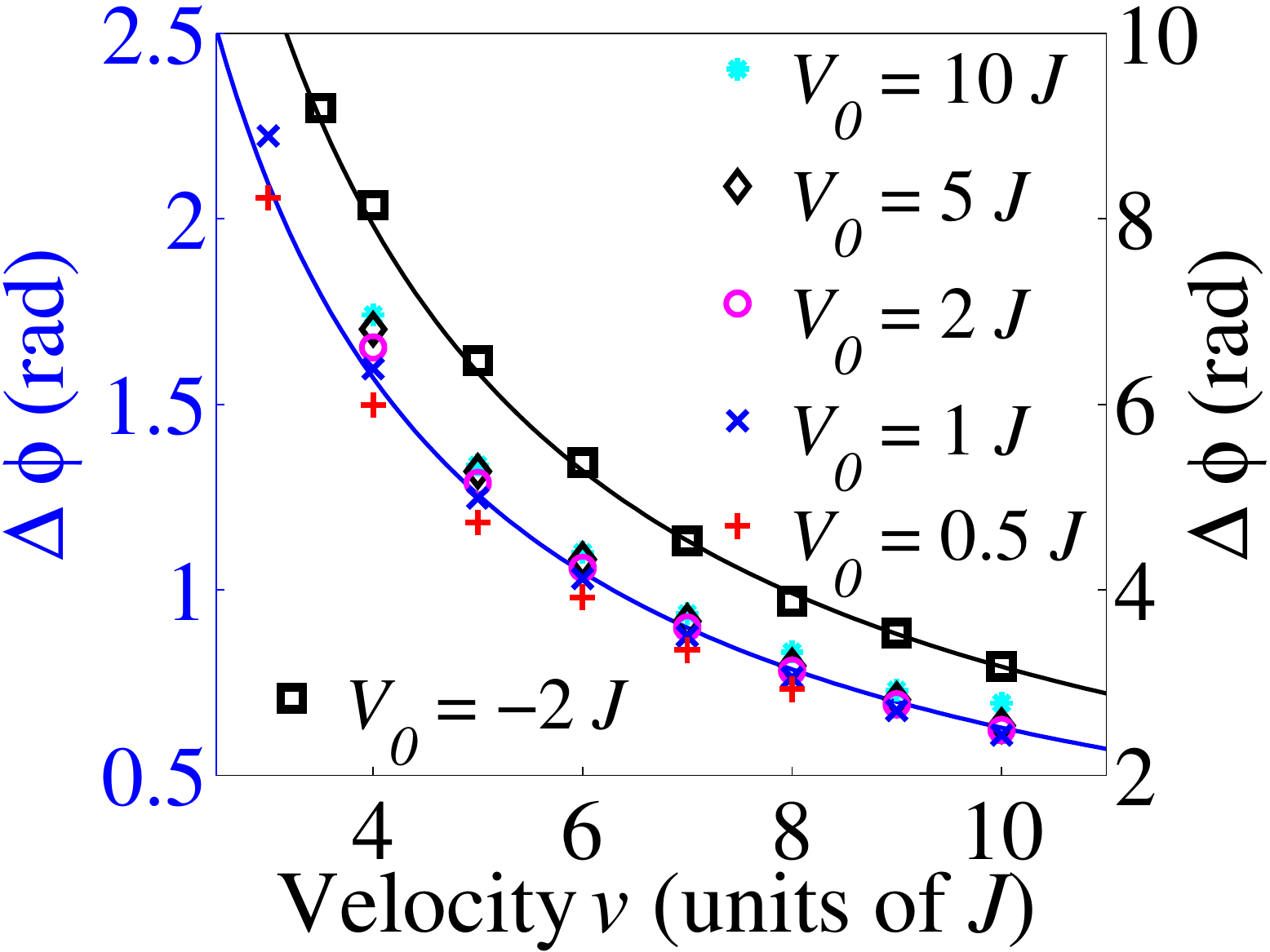}}}$
\caption{(Color online). Left: The phase difference $\phi_{x, -x}$ for different values of $U$ and $v$. Right: The maximum of $\phi_{ij}$ for different velocities of the perturbation given by the many-body simulations. The figure includes data for a Gaussian well with $V_0 = -2 J$ (right $y$-axis) and Lorentzian barriers with different heights from $V_0 = 10 J$ to $V_0 = 0.5 J$ (left $y$-axis). The black line is the result of the time-integral for the Gaussian, $\Delta\phi(v) = 2 V_0 \sigma \frac{\sqrt{2 \pi}}{v}$, and the blue line for the Lorentzian, $\Delta\phi(v) = \frac{2 \pi}{v}$. The time-integral of the Lorentzian is independent of the height $V_0$ since its width is $\frac{1}{V_0}$.
}
\label{fig:curves}
\end{center}
\end{figure}

In the ground state, the pair correlation function $C_{ij}$ is a real quantity, but perturbing the system gives it a nonzero time-dependent phase $\phi_{ij}(t)$,
\begin{align}
 \phi_{ij}(t) = \arctan{\left(\frac{\text{Im}[C_{ij}(t)]}{\text{Re}[C_{ij}(t)]}\right)}.
\end{align}
As the perturbation moves through the lattice, $\phi_{ij}$ changes across the perturbation center, as shown in Fig.~\ref{fig:pair_correlation_phase}. If one of the lattice site indices is fixed, for instance $i = 40$ in Fig~\ref{fig:pair_correlation_phase} b), and $\phi_{ij}$ observed at each site $j$, it can be seen to change smoothly from zero to approximately $\frac{5}{2} \pi$ when $j$ crosses the perturbation center. Similarly, by fixing $j = 40$ and varying $i$ one sees that the phase of $C_{ij}^* = C_{ji}$ changes from zero to approximately $-\frac{5}{2} \pi$. The value stays constant over a long range, i.e. up to very small values of the power-law decaying $|C_{ij}|$, which indicates a high numerical stability of the calculations. In the non-interacting case, the phase is not equally smooth and the density is more deformed (see Supplemental Material) \cite{supplemental}. In the case of a slow perturbation, the phase is randomized due to the movement and localization of the doublons. On the left side of Fig.~\ref{fig:curves}, $\phi_{x,-x}$ is plotted at the time step when the perturbation is at the middle of the lattice. For $v = 4 \: J$ and $v = 3.5 \: J$, a stronger interaction $U = -10 \: J$ is included, which shows that the phase difference does not depend significantly on the interaction. This is because $v > v_{\text{g}}^{\text{max}}$ for both interactions. In the case of a Lorentzian barrier, the change in the phase is steeper due to the narrower shape of the potential and from positive to negative due to the opposite sign.

The maximum phase changes calculated for different velocities of the well and barrier are gathered in Fig.~\ref{fig:curves} (right). The velocities are in the fast regime where the pair correlations are preserved. If the many-body system can to some extent be described by a single (macroscopic) wave function, the phase change can, in an extremely simplified model, be quantified by single-doublon dynamics. The wave function of a doublon can be written in the basis of localized states $\ket{\psi^{\text{d}}(t)} = \sum_{i = 0}^{L - 1}\alpha_i(t) \ket{d_i}$, where $\ket{d_i} = \ket{0, 0, \cdots, (\uparrow\downarrow)_i, 0, \cdots, 0}$. The time-dependent term $H_V(t)$ of eq.~(\ref{eq:hamiltonian_te}) does not commute with the kinetic term in $H_0$, but since the particles are only slightly displaced in the fast velocity regime, the kinetic term can be neglected, leaving $\tilde{H}_0 = H_U + H_{\text{trap}}$ and $\tilde{H}(t) = \tilde{H}_0 + H_V(t)$. The time evolution of the wave function is given by
\begin{align*}
  \ket{\psi^{\text{d}}(t)} &\approx e^{-i \int_0^t \tilde{H}(\tau) d\tau} \ket{\psi^{\text{d}}(0)} \\
  &= \sum_i e^{- 2i \int_0^t V(i, \tau) d\tau} e^{-i \tilde{H}_0 t} \alpha_i(0) \ket{d_i}.
\end{align*}
The factor of 2 in the exponent comes from the sum over $\sigma$ in $H_V(t)$. Considering a time $t$ when the narrow perturbation has passed the site $i$, the time evolution of another far-away site $j$ is given by $e^{-i \tilde{H}_0 t}$, and relative to $\alpha_j(t)$, $\alpha_i(t)$ has gathered a phase $\Delta\phi = 2 \int_{0}^t V(i, \tau) d\tau$. The pair correlation is $\bra{\psi^{\text{d}}} c_{i\uparrow}^{\dagger} c_{i\downarrow}^{\dagger} c_{j\downarrow} c_{j\uparrow} \ket{\psi^{\text{d}}} = e^{-i \Delta \phi}|\alpha_i||\alpha_j|$ for this single-doublon state. The integral that gives $\Delta\phi$ does not depend on $i$. Since the functions $V(i, \tau)$ decay quickly, the integration limits can be extended to $\pm \infty$ in order to obtain analytical expressions for $\Delta\phi$. They can be compared to the values of $\phi_{ij}$ obtained from the many-body simulations. The data points in Fig.~\ref{fig:curves} are the maxima of $\phi_{ij}$ over the lattice, and the curves are the results of $2 \int_{-\infty}^{\infty} V(i, \tau) d\tau$. The simple model describes the data remarkably well.

In conclusion, our results constitute one more striking demonstration of the peculiar nature of 1D physics compared to higher dimensions. Slow perturbations can break the initial pair correlations due to the existence of charge excitations at low energies around $q = 2k_F$. For such an excitation spectrum, the critical velocity in the sense of Landau's criterion would be zero. In the fast regime, the doublons do not have enough time to move and localize. Since the particle-hole spectrum in a lattice has an upper limit on energy, the fast perturbation can be interpreted as probing the high-velocity area where there are no states available. Correlations are preserved and a phase is imprinted on the 1D superfluid. Our predictions can be tested in state-of-the-art experiments with ultracold gases in optical lattices since the temperatures in lattice Fermi gases \cite{Esslinger} are already close to those where 1D superfluid correlations are predicted \cite{Batrouni, Heikkinen}. Phase imprinting in Fermi gases has been realized with a static laser beam \cite{Zwierlein}, and an interesting question is whether a situation similar to the fast perturbation studied here could be achieved in higher dimensions if the geometry of the perturbation was changed accordingly, e.g. a sheet moving through a 2D system. 

\begin{acknowledgments}
We thank T. Giamarchi for useful discussions. This work was supported by the Academy of Finland through its Centres of Excellence Programme (projects No. 139514, No. 251748, No. 135000, No. 141039, No. 263347 and
No. 272490) and by the European Research Council (ERC-2013-AdG-340748-CODE). A.-M. V. acknowledges financial support from the Vilho, Yrj\"o and Kalle V\"ais\"al\"a Foundation. D.-H. K. acknowledges support from Basic Science Research Program through the National Research Foundation of Korea funded by the Ministry of Science, ICT \& Future Planning (NRF-2014R1A1A1002682) and GIST college's GUP research fund. F. M. acknowledges financial support from the ERC Advanced Grant MPOES. Computing resources were provided by CSC--the Finnish IT Centre for Science 
 and the Aalto Science-IT Project. 
\end{acknowledgments}

\end{document}